\documentclass[final,5p,times,twocolumn]{elsarticle}

\usepackage{amssymb}
\usepackage{amsmath}
\usepackage{lipsum}
\usepackage[utf8]{inputenc}
\usepackage[T1]{fontenc}
\usepackage[spanish,galician,catalan,english]{babel}
\usepackage{todonotes}
\usepackage{hyperref}

\newcommand{\jpsi}{J/\psi}
\newcommand{\ups}{\Upsilon(1S)}
\newcommand{\mD}{m_D}
\newcommand{\Tc}{T_c}
\newcommand{\ReV}{\mathrm{Re}\,V}
\newcommand{\ImV}{\mathrm{Im}\,V}
\newcommand{\raa}{R_{AA}}
\newcommand{\sqrtsNN}{\sqrt{s_{NN}}}
\newcommand{\pT}{p_T}

\journal{Physics Letters B}

\begin{document}

\begin{frontmatter}

\title{Lindblad-driven quarkonium production in heavy-ion collisions}

\author[first]{N\'estor Armesto}
\author[second]{Miguel Ángel Escobedo}
\author[first]{Elena G. Ferreiro}
\author[first]{Víctor López-Pardo}
\affiliation[first]{organization={Instituto Galego de Física de Altas Enerxías IGFAE, Universidade de Santiago de Compostela},
            addressline={Rúa de Xoaquín Díaz de Rábago, s/n}, 
            city={Santiago de Compostela},
            postcode={15782}, 
            state={Galicia},
            country={Spain}}
\affiliation[second]{organization={Departament de Física Quàntica i Astrofísica and Institut de Ciències del Cosmos, Universitat de Barcelona},
            addressline={Martí i Franquès 1}, 
            city={Barcelona},
            postcode={08028}, 
            state={Catalonia},
            country={Spain}}


\begin{abstract}
We study the production of the conventional quarkonium states in ultrarelativistic heavy-ion collisions using an open quantum system
framework based on the Lindblad equation.  Starting from the complex-valued
in-medium potential,
we derive the
dissociation temperature and thermal decay width for each state, and compute their
survival probabilities for a system undergoing Bjorken expansion.
We then extend the framework to
include recombination from thermalized charm and bottom quarks in the
quark-gluon plasma, deriving a coalescence model {for quarkonia} from the Lindblad equation
under the adiabatic approximation.  
The methodology provides a unified, first-principles-inspired description of
suppression and recombination for both charmonium and bottomonium.
\end{abstract}


\end{frontmatter}



\section{Introduction}
\label{sec:intro}

Heavy quarkonium states have long served as precision probes of the quark-gluon plasma (QGP)
formed in ultrarelativistic heavy-ion collisions~\cite{Matsui:1986dk,
Andronic:2015wma}.  The interplay of Debye screening of the
heavy-quark potential and in-medium decoherence leads to a suppression of
quarkonium yields that is ordered hierarchically with binding energy: the more
weakly bound the state, the lower the temperature at which it dissolves in the
plasma.

On the theoretical side, a significant advance has been the recognition that the
in-medium heavy-quark potential is a complex-valued quantity~\cite{Laine:2006ns,Beraudo:2007ky}: the real part exhibits Debye screening while
the imaginary part encodes Landau damping of the mediating gluons. A systematic
and phenomenologically successful parametrization of this potential is provided by
the Gauss law model of Lafferty and Rothkopf~\cite{Lafferty:2019jpr}, which
combines the Cornell potential in vacuum with Hard-Thermal-Loop (HTL)
perturbation theory for the medium response, reproducing nonperturbative lattice
QCD results for both $\ReV$ and $\ImV$ with a single temperature-dependent
parameter, the Debye mass $\mD$.

A complementary line of development treats quarkonium as an open quantum system
coupled to the QGP bath~\cite{Akamatsu:2012vn,Brambilla:2016wzb,
Blaizot:2017ypk,Brambilla:2019tya}.  In this formalism the evolution of the
density matrix of the heavy-quark pair is governed by a Lindblad equation, and
the survival probability of a pre-formed bound state can be expressed in terms of
the imaginary part of the complex potential.  More recently, we have shown that the same framework also provides a natural coalescence model obtained by projecting the stochastic jumps of the Lindblad
equation onto the bound-state subspace~\cite{Armesto:2024xmm,Armesto:2025lbd}.
Applied to the exotic state $X(3872)$, assumed to be a compact tetraquark, this
approach predicted a sizable enhancement of the nuclear modification factor $\raa$
arising from recombination, compatible with CMS observations~\cite{Sirunyan:2021qef}.

The goal of the present work is to apply this unified suppression-plus-recombination framework
to the conventional quarkonium ground states.
The paper is organized as follows.  In Section~\ref{sec:potential} we recall the
Gauss law complex potential and its parametrization.
In Section~\ref{sec:spectral} we derive the in-medium spectral functions for charmonium and bottomonium and we extract the binding energies, the thermal decay widths and
the dissociation temperatures for $\jpsi$, $\psi(2S)$ and $\Upsilon(nS)$.
In Section~\ref{sec:survival} we present the survival probability under Bjorken
expansion derived within the recombination model from the
Lindblad equation.
Section~\ref{sec:phenomology} collects the phenomenological inputs and presents
results for $\raa$.  Section~\ref{sec:summary} contains our conclusions and
outlook.

\section{In-medium potential}
\label{sec:potential}


The starting point is the well-established Cornell form of
the vacuum heavy-quark potential \cite{Eichten:1978tg}:
\begin{equation}
  V_\text{vac}(r) = -\frac{\tilde\alpha_s}{r} + \sigma\, r + c \,,
  \label{eq:cornell}
\end{equation}
where $\tilde\alpha_s = C_F g^2/(4\pi)$ is the strong coupling including the
Casimir factor $C_F$, $\sigma$ is the string tension, and $c$ is an additive
renormalization constant.  This potential captures both asymptotic freedom at
short distances and linear confinement at large distances.

The parameters $\tilde\alpha_s$, $\sqrt{\sigma}$, and $c$ could be fixed independently
for the charmonium and bottomonium systems but, {assuming} the heavy quark potential is universal {since} at lowest order in pNRQCD
the same expressions arise for both heavy quark families, we will fix the values for bottomonium
and tune the charm mass to reproduce the $\jpsi$ and $\psi(2S)$ masses. The renormalon-subtracted bottom quark mass
$m_b^{RS'} = 4.882\;\text{GeV}$~\cite{Pineda:2001ra} is used for the
bottomonium sector, while the charm mass $m_c^\text{fit}=1.4692\;\text{GeV}$ reproduces
the $\jpsi$ and $\psi(2S)$ masses. The resulting parameters, common to
Ref.~\cite{Lafferty:2019jpr}, are: $\tilde\alpha_s=0.513 \pm 0.002$, $\sqrt{\sigma}=0.412 \pm 0.004$ GeV and $c=-0.161 \pm 0.003$ GeV. 

At finite temperature the vacuum potential is modified by the QGP medium. 
The in-medium potential is written as
\begin{equation}
V(r,T) = \mathrm{Re}\,V(r,m_D(T)) + i\,\mathrm{Im}\,V(r,m_D(T)),
\label{eq:potential}
\end{equation}
where $r$ is the heavy quark-antiquark separation and $m_D(T)$ the Debye screening mass that encodes
color screening in the QGP.  

The real part of the in-medium potential is obtained by applying a linear-response
(polarization/permittivity) procedure to the vacuum 
potential. Using the HTL permittivity \(\varepsilon(p,m_D)\)
according to \(V(p)=V_{\mathrm{vac}}(p)/\varepsilon(p,m_D)\) and enforcing smooth
matching to the vacuum potential for \(m_D\to 0\), one obtains
\begin{equation}
\begin{aligned}
    \mathrm{Re}\,V(r,m_D(T))&= -\tilde{\alpha}_s\left(m_D+\frac{e^{-m_Dr}}{r}\right)+c\\
&\quad+\sigma\left[\frac{2}{m_D}(1-e^{-m_Dr})-re^{-m_Dr}\right],
\label{ReVJY}
    \end{aligned}
\end{equation}
which reproduces the HTL result for the Coulomb term and the {known} string-term expression~{\cite{Lafferty:2019jpr}}.

The imaginary part of the potential arises from Landau damping and the scattering of the
heavy quark with medium gluons, which induces a finite in-medium width. 
While the Coulomb term of the imaginary part matches the
HTL result, the string term needs to be regularized. Following~\cite{Lafferty:2019jpr} it is possible to rigorously remove the nonphysical divergence. The imaginary part of the quarkonium potential can then be written as
\begin{equation}
    \mathrm{Im}\,V(r,m_D(T))=-\tilde{\alpha}_sT\phi(m_Dr)+\frac{\sigma T}{m_D^2}\chi(m_Dr,\Delta_D)~,\label{ImVJY}
\end{equation}
with
\begin{equation}
    \chi(x,\Delta_D)=2\int_0^\infty du\ \frac{2-2\cos{ux}-ux\sin{ux}}{\sqrt{u^2+\Delta_D^2}(u^2+1)^2}~.
\end{equation}
Choosing the regularization constant to be $\Delta_D\simeq3.0369$, the string term has a similar behavior to the Coulomb term at large distances since $\lim_{x\to\infty}\chi(x,\Delta_D)=1$.
Thanks to the regularization, the imaginary part of the potential is constant at large
distances and high temperatures. This {imaginary} part is related to the decay width.

\begin{figure}[h]
    \centering
    \includegraphics[scale=0.45]{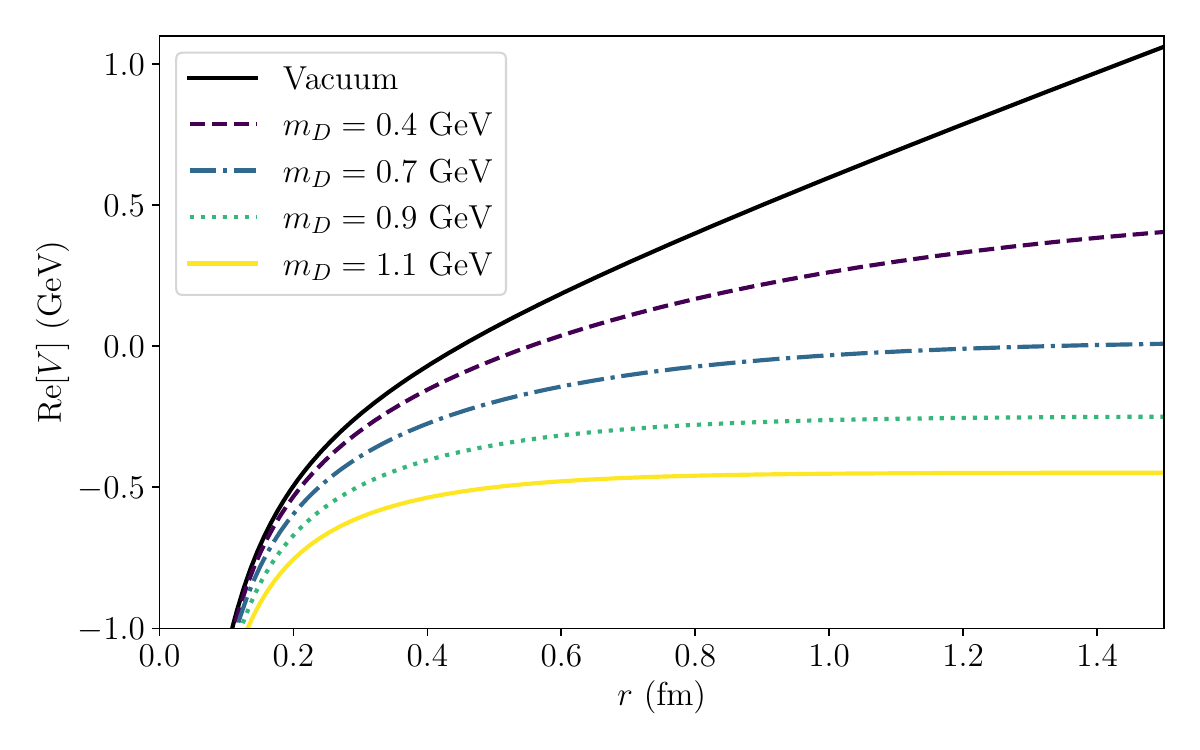}
    \caption{Real part of the in-medium potential for quarkonia as given by~\eqref{ReVJY}. It is clear that the increase in temperature (and Debye mass) makes the potential barrier smaller.}
    \label{fig:ReVCornell}
\end{figure}

\begin{figure}[h]
    \centering
    \includegraphics[scale=0.45]{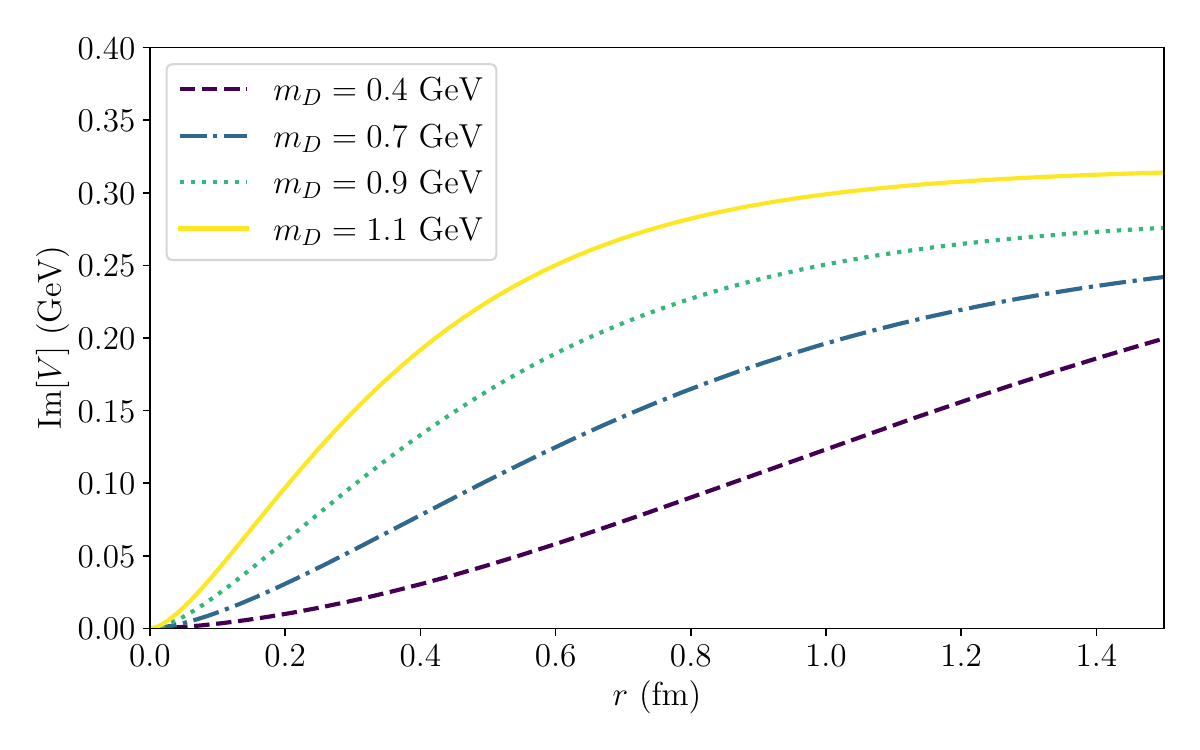}
    \caption{Imaginary part of the in-medium potential for quarkonia as given by~\eqref{ImVJY}. Only finite temperature cases are shown since the vacuum potential is purely real.}
    \label{fig:ImVCornell}
\end{figure}

The behaviour of both the real and imaginary parts of the potential are shown in Figures~\ref{fig:ReVCornell} and \ref{fig:ImVCornell}.


\section{In-medium spectral functions and dissociation properties}
\label{sec:spectral}


In the potential model approach, the in-medium quarkonium eigenenergies can be
obtained by solving the Schrödinger equation with the complex potential. However, for a potential with position-dependent imaginary part, the time-space solution of the Schrödinger equation becomes challenging since schemes used for Hermitian Hamiltonians, like Numerov's method used in~\cite{Armesto:2024xmm,Armesto:2025lbd}, become inapplicable. Nevertheless, the diagonalization of the Hamiltonian is possible and simpler in the frequency domain. The in-medium spectral functions are obtained by solving the frequency-space Schrödinger equation with the complex potential and extracting the imaginary part of the corresponding Green’s function, which directly yields the quarkonium spectral shape, including both bound-state peaks and their thermal broadening~\cite{Burnier:2007qm}.

Specifically, the spectral function is obtained as the imaginary part 
of the traced retarded Green's function of the Schr\"odinger equation 
with the complex potential,
\begin{equation}
\rho(\omega) = -\frac{1}{\pi} \,\mathrm{Im} \int d^3r\, G^R(r,r;\omega),
\end{equation}
where $G^R(r,r';\omega) = \langle r | (\omega + i\epsilon - H)^{-1} | r' \rangle$ 
and $H$ contains the complex in-medium potential $V(r,T)$.

In the vicinity of each peak, {when the decay width is a perturbation,} the spectral function is well approximated by a Breit--Wigner distribution,
\begin{equation}
\rho(\omega) \approx \frac{1}{\pi}\frac{\Gamma_n/2}{(\omega - E_n)^2 + (\Gamma_n/2)^2},
\end{equation}
where $E_n$ is the peak position and $\Gamma_n$ is the full width at half maximum. In practice and following~\cite{Lafferty:2019jpr}, the binding energy of each state is determined by identifying $E_n$ with the in-medium mass and subtracting the sum of the constituent quark masses, while the thermal decay width is read off directly as the width of the peak measured at half of its maximum height. Since the spectral peaks are in general asymmetric, we do not fit a symmetric 
Breit-Wigner directly to the numerical spectral function. Instead, we employ 
a skewed Breit-Wigner profile, which better captures the distortion of the 
line shape near the continuum threshold and allows for a more reliable 
extraction of both $E_n$ and $\Gamma_n$ across the full temperature range.

\subsection{Charmonium}
\begin{figure}[h]
    \centering
    \includegraphics[width=1\linewidth]{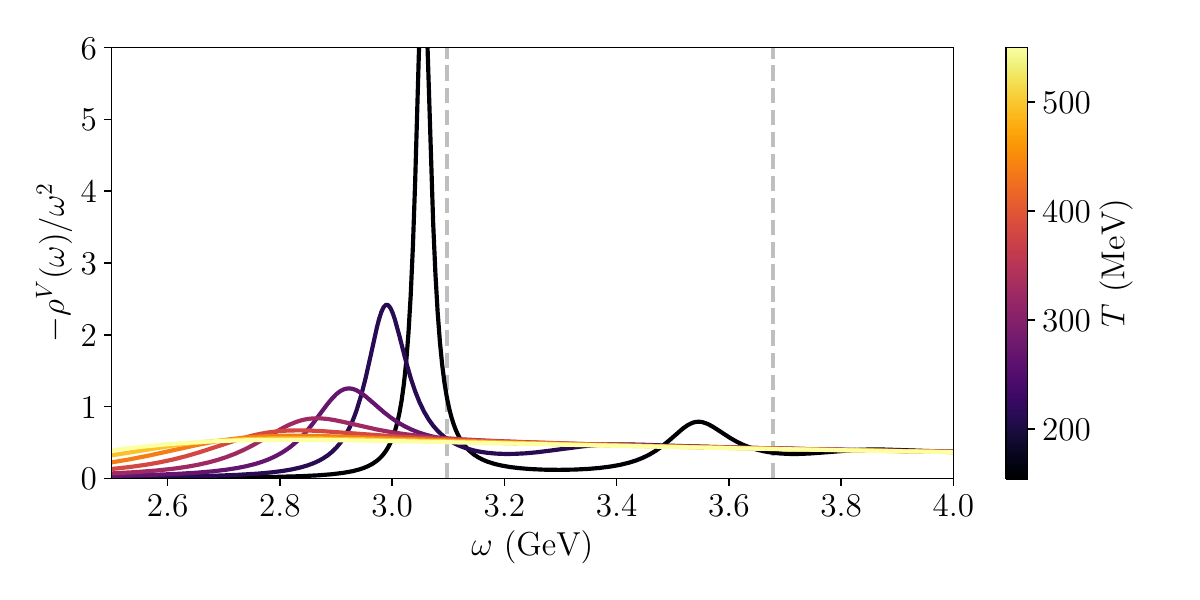}
    \caption{In-medium $S$-wave spectral functions for vector channel charmonium for different temperatures. The dashed gray vertical lines represent the $T=0$ bound states $J/\psi$ and $\psi(2S)$.}
    \label{fig:CharmSpectrum}
\end{figure}

The charmonium spectral function at various temperatures, displayed in Figure~\ref{fig:CharmSpectrum}, exhibits two peaks at finite temperature. 

As the temperature increases, these peaks gradually shift toward lower frequencies and eventually fade away. At the same time, they become broader and less pronounced, indicating a shorter lifetime of the states. In the vacuum, the spectral function reduces to two Dirac delta functions located at the masses of the $\jpsi$ and $\psi(2S)$, shown as dashed gray lines in the figure. By fitting the regions around the peaks with a skewed Breit–Wigner function one can extract how both the binding energy and decay width depend on temperature. 
The binding energy is determined from the peak position, interpreted as the in-medium mass, after subtracting the masses of the two constituent quarks. The decay width is given by the width of the peak measured at half of its maximum height.

\begin{figure}[h]
    \centering
    \includegraphics[scale=0.45]{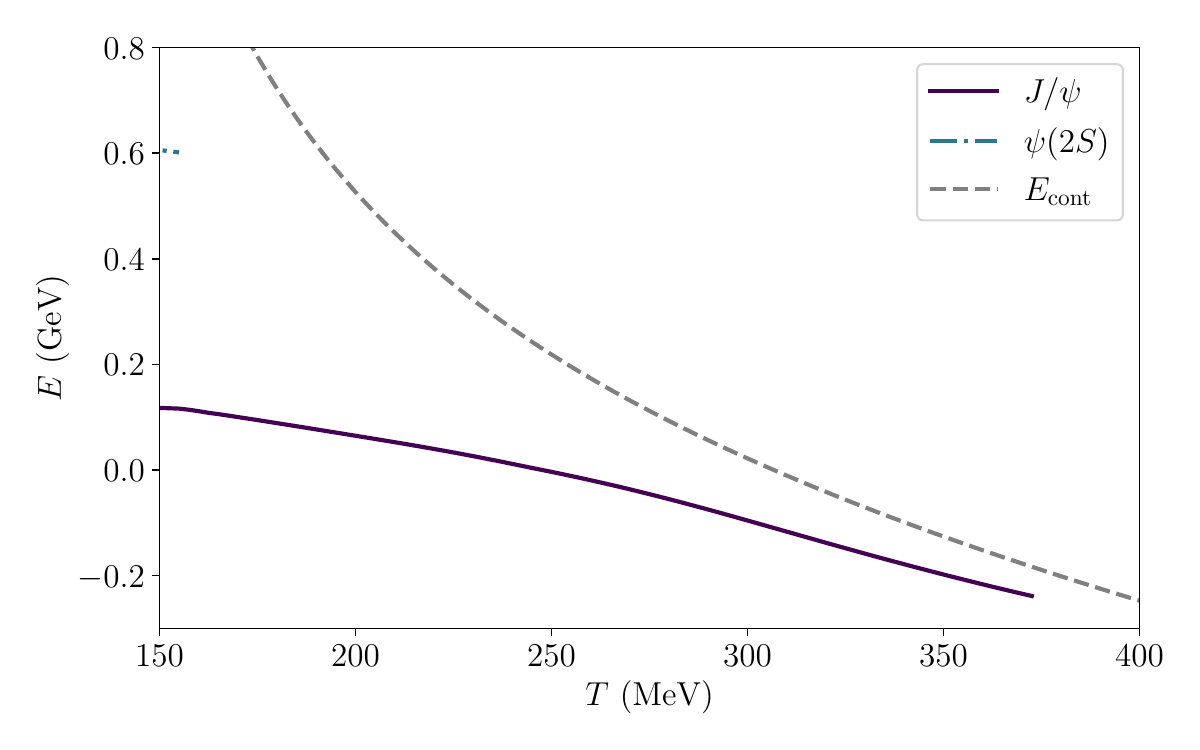}
    \caption{Binding energy for the different species of charmonium at finite temperature: $J/\psi$ (solid purple line) and $\psi(2S)$ (dash-dot blue line). The potential barrier is also shown  (dashed gray line). The $\psi(2S)$ line can barely be seen because the corresponding peak disappears at $T\gtrsim155\ \rm MeV$.}
    \label{fig:CharmEnergy}
\end{figure}

The binding energy obtained with this method is shown in Figure~\ref{fig:CharmEnergy}. From this figure, it is evident that the binding energy of the $J/\psi$ state can be tracked over nearly the entire temperature range, whereas the corresponding signal for the $\psi(2S)$ state is barely visible. This is due to the fact that the fit parameters of the $\psi(2S)$ spectral function cannot be reliably extracted for temperatures $T \gtrsim 155~\text{MeV}$. As observed in Figure~\ref{fig:CharmSpectrum}, the $\psi(2S)$ peak essentially disappears for all temperatures except the lowest one.
In contrast, the binding energy of the $J/\psi$ decreases steadily with increasing temperature, in agreement with the mass shift observed in Figure~\ref{fig:CharmSpectrum}.

\begin{figure}[h]
    \centering
    \includegraphics[scale=0.45]{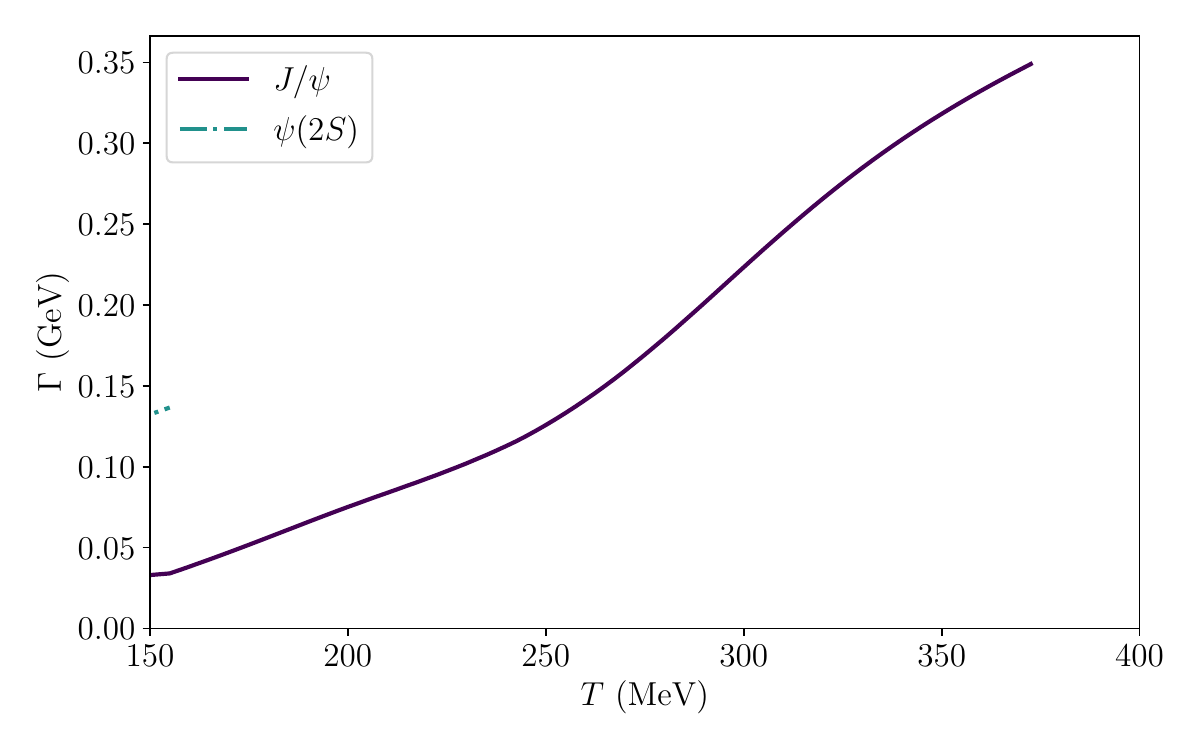}
    \caption{Decay width for the different species of charmonium at finite temperature: $J/\psi$ (solid purple line) and $\psi(2S)$ (dash-dot blue line). The $\psi(2S)$ line is barely visible because the corresponding peak vanishes at $T\gtrsim155\ \rm MeV$.}
    \label{fig:CharmGamma}
\end{figure}

The widths of the peaks shown in Figure~\ref{fig:CharmSpectrum} correspond to the decay widths of the charmonium states and are presented in Figure~\ref{fig:CharmGamma}. As in the case of the binding energy, the $\psi(2S)$ peak is difficult to resolve, and therefore its decay width can only be reliably extracted for temperatures below $155~\text{MeV}$. 
On the other hand, the $J/\psi$ peak remains well defined over a much broader temperature range, allowing its decay width to be determined across a wide interval.
Eventually, the $J/\psi$ state dissociates, meaning that the corresponding peak disappears. Beyond this point, the notion of a decay width loses its meaning, and consequently it is no longer displayed in Figure~\ref{fig:CharmGamma}.

\textcolor{violet}{}

The dissociation temperature $T_d$ is defined as the temperature above which the Schr\"odinger equation no longer admits a bound-state solution, meaning that the corresponding spectral peak merges into the continuum. Equivalently, $T_d$ can be identified as the temperature at which the in-medium binding energy vanishes.
In practice, the dissociation temperature can be estimated from the spectral function as the highest temperature at which a distinct peak is still visible in Figure~\ref{fig:CharmSpectrum}. This corresponds, in turn, to the last temperature point at which the given charmonium state appears in Figures~\ref{fig:CharmEnergy} and \ref{fig:CharmGamma}. 
Using this approach, the dissociation temperatures of charmonium states are found to be approximately
\[
T_{J/\psi} \approx 372~\text{MeV}, \qquad T_{\psi(2S)} \lesssim 155~\text{MeV}.
\]
These values are physically reasonable: the $\psi(2S)$, being weakly bound, dissociates at relatively low temperatures and barely survives in the medium, whereas the more tightly bound $J/\psi$ persists up to higher temperatures.

\subsection{Bottomonium}
\begin{figure}[h]
    \centering
    \includegraphics[width=1\linewidth]{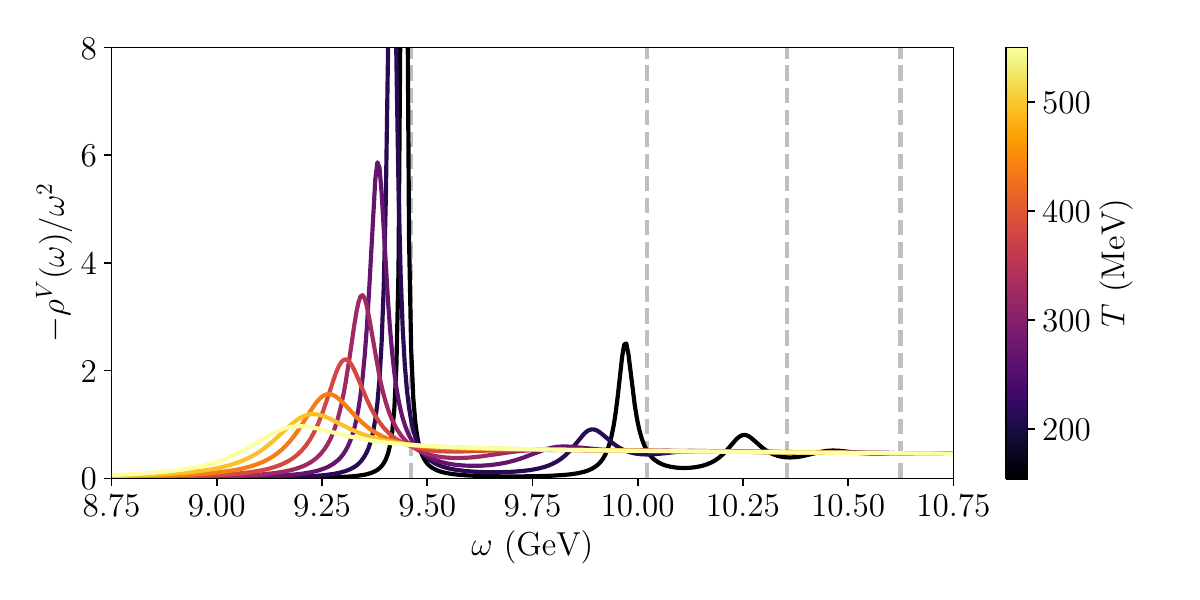}
    \caption{In-medium $S$-wave spectral functions for vector channel bottomonium for different temperatures. The dashed gray vertical lines represent the $T=0$ bound states: $\Upsilon(1S)$, $\Upsilon(2S)$, $\Upsilon(3S)$ and $\Upsilon(4S)$.}
    \label{fig:BottomSpectrum}
\end{figure}

The bottomonium spectral function is obtained following the same procedure as in the charmonium case. As shown in Figure~\ref{fig:BottomSpectrum}, increasing the temperature leads to a shift of the bottomonium peaks towards lower frequencies, accompanied by a broadening. The dashed gray lines indicate the Dirac delta functions corresponding to the spectral function in the vacuum. Although four $S$-wave states are present at zero temperature, only three of them remain at finite temperature.
As in the charmonium case, the spectral function in the vicinity of each peak can be fitted with a skewed Breit--Wigner distribution, allowing for the extraction of both the binding energy and the decay width. For bottomonium, the first two peaks are much more clearly resolved than in the charmonium case, enabling more precise fits, particularly at low temperatures. In contrast, the $\Upsilon(4S)$ state does not survive in the medium, and its corresponding peak is only present in the vacuum.

\begin{figure}[h]
    \centering
    \includegraphics[scale=0.45]{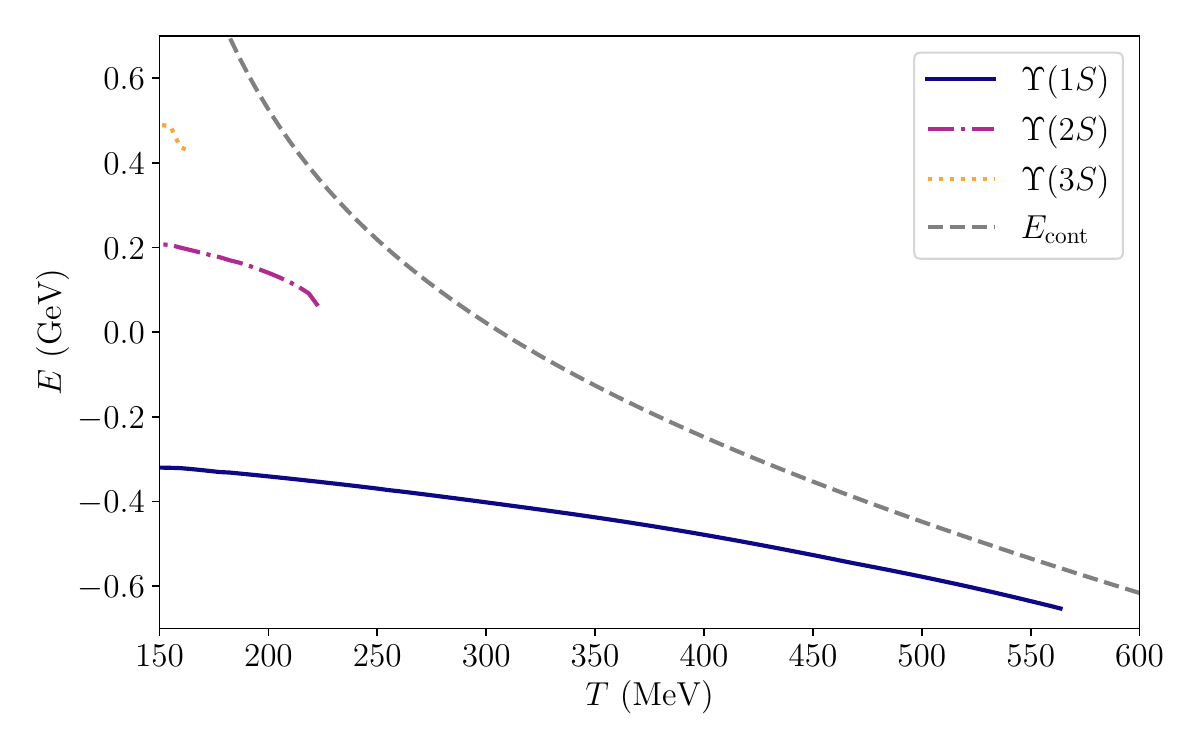}
    \caption{Binding energy for the different species of bottomonium at finite temperature: $\Upsilon(1S)$ (solid blue line), $\Upsilon(2S)$ (dash-dot magenta line), and $\Upsilon(3S)$ (dotted orange line). The potential barrier is also shown (dashed gray line).}
    \label{fig:BottomEnergy}
\end{figure}

The binding energy, obtained from the position of the peak, is shown in Figure~\ref{fig:BottomEnergy}. Similarly to charmonium, the binding energy of bottomonium states decreases with increasing temperature. As expected, more weakly bound states dissociate at lower temperatures: the $\Upsilon(1S)$ survives up to higher temperatures than the $\Upsilon(2S)$, which in turn persists longer than the $\Upsilon(3S)$. This hierarchy is evident in Figure~\ref{fig:BottomEnergy}, where states with larger initial binding energies extend over a wider temperature range. As the binding energy approaches the potential barrier, the corresponding peak becomes unresolvable, signaling the onset of dissociation.

\begin{figure}[h]
    \centering
    \includegraphics[scale=0.45]{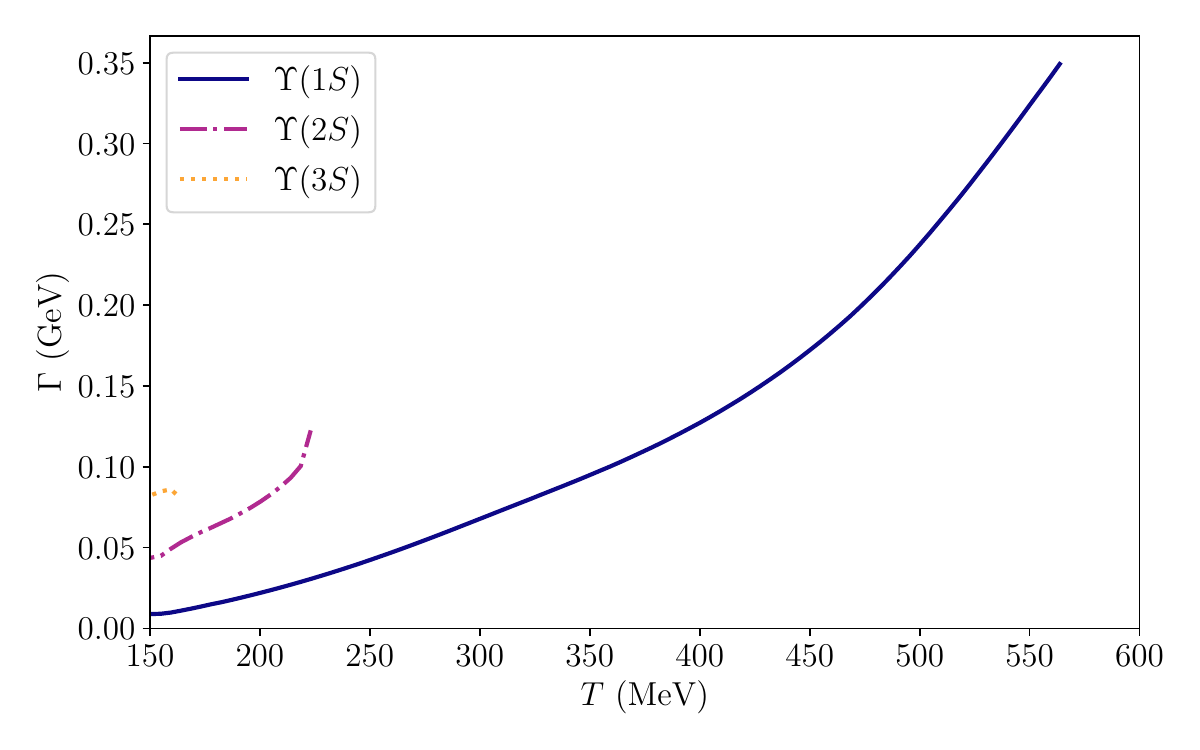}
    \caption{Decay width for the different species of bottomonium at finite temperature: $\Upsilon(1S)$ (solid blue line), $\Upsilon(2S)$ (dash-dot magenta line), and $\Upsilon(3S)$ (dotted orange line).}
    \label{fig:BottomGamma}
\end{figure}

Moreover, increasing the temperature leads to a broadening of the peaks, corresponding to an increase in the decay width. The decay width is therefore a monotonically increasing function of temperature, as shown in Figure~\ref{fig:BottomGamma}.

From this analysis, one can estimate the dissociation temperatures of the different bottomonium states. Ordered from most to least tightly bound, they are approximately
\[
T_{\Upsilon(1S)} \approx 564~\text{MeV}, \,
T_{\Upsilon(2S)} \approx 223~\text{MeV}, \,
T_{\Upsilon(3S)} \approx 164~\text{MeV}.
\]
The $\Upsilon(4S)$ state dissociates at temperatures below the critical temperature, $T_c \approx 155~\text{MeV}$, and therefore does not form within the medium. This provides an upper bound for its dissociation temperature, $T_{\Upsilon(4S)} < 155~\text{MeV}$.

\section{Survival probability}
\label{sec:survival}

\subsection{Suppression-only scenario}

For a quarkonium state produced at time $t_0$ and traversing a QGP that cools
according to Bjorken hydrodynamics~\cite{Bjorken:1982qr}, $T(t) = T_0\,(t_0/t)^{1/3}$,
the survival probability can easily be determined through the decay width 
\begin{equation}
  S(t_0,t) = \exp\!\left[-\int_{t_0}^{t} d\tau\;\Gamma(T(\tau))\right] .
  \label{eq:survival}
\end{equation}
If the initial temperature $T_0 > T_d$, the state is not formed; otherwise
Eq.~\eqref{eq:survival} applies from the formation time $t_0$ until the
freeze-out time $t_f$ defined by $T(t_f) = \Tc$, 
around the phase transition value.

\subsection{Lindblad framework and recombination}
\label{sec:lindblad}

The inclusion of recombination is modeled by assuming that heavy quarks move non-relativistically around their center of mass and that $\Lambda_{\text{QCD}}$ is larger than the binding energy. 
At finite temperature, the potential is modified by the medium. However, since the in-medium potential only captures the time-ordered correlator of the heavy quark--antiquark pair, a complete description of bound-state populations requires treating quarkonium as an open quantum system and following the evolution of its density matrix.

In the case $T\gg E$, where $T$ is the temperature of the medium and $E$ corresponds
to the binding energy of the state that would conform the system, the evolution of
the density matrix $\rho$ is given by a Lindblad equation~\cite{Akamatsu:2012vn,Blaizot:2017ypk,Brambilla:2016wzb,Brambilla:2019tya}:
\begin{equation}
  \frac{d\rho}{dt} = -i[H,\rho]
  + \sum_n\!\left(C_n\rho C_n^\dagger
    - \tfrac{1}{2}\{C_n^\dagger C_n,\rho\}\right) ,
  \label{eq:lindblad}
\end{equation}
where $H$ is the Hamiltonian of the $Q\bar{Q}$ pair and the $C_n$ are collapse
operators describing transitions between color configurations and interactions
with medium gluons.

The associated effective non-Hermitian Hamiltonian is
\begin{equation}
  H_\text{eff} = H - \frac{i}{2}\sum_n C_n^\dagger C_n  \,,
  \label{eq:heff}
\end{equation}
where 
  $\ImV(r) = -\frac{1}{2}\sum_n C_n^\dagger C_n$ \,.
With this redefinitions, the Lindblad equation can be written as
\begin{equation}
      \frac{d\rho}{dt} =-iH_{\mathrm{eff}}\rho+i\rho H_{\mathrm{eff}}^\dagger+\sum_nC_n\rho C_n^\dagger~.\label{effectiveLindblad}
\end{equation}

Note that the development done in the previous section, focused on the Schrödinger equation with a non-Hermitian Hamiltonian, is equivalent to neglecting the last term on~\eqref{effectiveLindblad}. This last term is the only one that takes into account the possibility of the regeneration of the bound-state wavefunction.

To compute {uncorrelated} recombination from the Lindblad equation~\cite{Armesto:2025lbd}, several assumptions are introduced. First, the heavy quark–antiquark system is described by a Markovian Lindblad equation, where the medium induces decoherence and transitions between different states. We work in the dilute heavy-quark limit, considering only binary interactions and neglecting higher-order correlations. Unbound heavy quarks are assumed to be uncorrelated and locally thermalized with the medium (molecular chaos). In addition, an adiabatic approximation is used, meaning that the effective Hamiltonian and jump operators evolve slowly compared to the intrinsic timescales of the bound states, which allows one to define survival probabilities and recombination rates. It is important to note that this derivation differs slightly from the one in~\cite{Armesto:2025lbd}. We make use of the non-Hermitian adiabatic theorem of~\cite{Kumar:2025quk}, which states that at large times any initial state gets converted to the state with the smallest decay width. Since we are studying the fundamental state and we expect that to be also the ``less decaying state'', the overall conclusion is the same; the bound state of the effective Hamiltonian at time $t$ is mapped to the bound state of the Hamiltonian at time $t'$.  Finally, recombination is assumed to be rare, so the population of unbound heavy quarks remains approximately constant during the evolution.


Following the derivation in~\cite{Armesto:2025lbd}, we project the Lindblad
equation onto the bound-state subspace $P_b = \sum_i |i\rangle\langle i|$ and
free subspace $P_f$.  The probability per unit time for a stochastic jump from a
free $Q\bar{Q}$ state to the bound state $|1S\rangle$ through channel $n$ is
\begin{equation}
  R_n = Z_{t,n}\,\mathrm{Tr}\!\left[\rho_{t,n}\,\rho_f\right] ,
  \label{eq:Rn}
\end{equation}
where the target density matrix and its normalization are
\begin{equation}
  \rho_{t,n} = \frac{C_n^\dagger P_b C_n}{Z_{t,n}} \,, \qquad
  Z_{t,n} = \mathrm{Tr}\!\left[C_n C_n^\dagger P_b\right] .
\end{equation}
Assuming that free heavy quarks are uncorrelated and thermalized (molecular chaos), their
Wigner distribution is
\begin{equation}
  W_f(\mathbf{R},\mathbf{r},\mathbf{P},\mathbf{p})
  = Z_f\,N^2\,e^{-\frac{P^2}{4MT} - \frac{p^2}{MT}} \,,
  \quad Z_f^{-1} = V^2\!\left(\frac{MT}{2\pi}\right)^3 ,
  \label{eq:Wf}
\end{equation}
where $M$ is the heavy-quark mass, $V$ the medium volume, and $N=N_Q=N_{\bar Q}$ the number of free heavy quarks.  Substituting into Eq.~\eqref{eq:Rn} and using the Wigner
transform leads to
\begin{equation}
  R_{1S}(t) = \frac{Z_{t,1S}\, N^2}{V
    \left(\frac{MT}{4\pi}\right)^{3/2}}
    \int \frac{d^3 p}{(2\pi)^3}\,e^{-\frac{p^2}{MT}}\,W_{t,1S}(\mathbf{p}) \,,
  \label{eq:R1S}
\end{equation}
where $W_{t,1S}(\mathbf{p})$ is the target Wigner distribution.  Following
Ref.~\cite{Armesto:2025lbd}, we approximate $W_{t,1S} \approx W_{1S}$, the Wigner
distribution of the bare bound-state wave function obtained by solving the Schrödinger equation in the previous section.
We note that in the high temperature limit the above integral equals~1. {We will use the high-temperature approximation from now on.} 


In Ref.~\cite{Armesto:2025lbd}, we have assumed that 
the normalization factor $Z_{t,1S}$ is equal to the thermal decay width of
the state  $Z_{t,1S}(T) = \Gamma_{1S}(T)$. This assumption relies on the fluctuation-dissipation theorem, which is only valid when the heavy quarks are fully thermalized. In fact, further corrections are needed for more realistic applications in heavy-ion collisions. In transport approaches~\cite{Grandchamp:2003uw}, this is typically handled by introducing a
relaxation factor which modulates the gain term in the rate equation,
\begin{equation}
    \mathcal{R}(t)=1-\exp\left(-\int_{t_0}^t d\tau/\tau_{Q}^{therm}\right),
    \label{eq:relfac}
\end{equation}
where $\tau_{Q}^{therm}$ represents the kinetic relaxation time of the heavy quark distributions, i.e., the time that the heavy quarks take to equilibrate within the medium. For charm quarks, it can be approximated by $\tau_{c}^{therm} \approx 4\;\text{fm}$~\cite{Song:2012at,Wu:2024gil}.
Taking the relaxation time as proportional to the quark mass, the corresponding bottom quark’s relaxation time is $\tau_{b}^{therm} \approx12$ fm.

Therefore, we define the high temperature limit of~\eqref{eq:R1S} multiplied by the relaxation factor~\eqref{eq:relfac},
\begin{equation}
  R(t) = \frac{Z_{t,1S}\, N^2}{V
    \left(\frac{MT}{4\pi}\right)^{3/2}}
    \mathcal{R}(t) \,,
  \label{eq:R}
\end{equation}
which is what we will use for phenomenological applications. We note that this is different from what we do for the $X(3872)$ in our previous paper~\cite{Armesto:2025lbd}, where the high temperature limit is only taken when made explicit and the relaxation factor never taken into account.

The contribution from coalescence to the probability for observing state $1S$ at
freeze-out time $t_f$ is, in the adiabatic approximation~\cite{Armesto:2025lbd,Messiah},
\begin{equation}
  N_{1S}^\text{recomb}(t_f)
  = \int_0^{t_f} dt\;S(t_f, t)\;R(t) \,,
  \label{eq:N_recomb}
\end{equation}
where $S(t_f, t)$ is the survival probability 
between times $t$ and $t_f$ computed from Eq.~\eqref{eq:survival}.  The adiabatic approximation is valid when the
bound-state energy levels change slowly compared to the time scale of a quantum
jump, i.e., when
\begin{equation}
  \frac{|\langle i | \dot{V} | j \rangle|^2}{(E_i - E_j)^4} \ll 1
  \qquad \forall\;i \neq j \,.
  \label{eq:adiabatic_condition}
\end{equation}
This inequality holds for both $\jpsi$ and $\ups$ across the temperature
range $[\Tc, T_d]$.

\section{Phenomenology in Pb--Pb collisions}
\label{sec:phenomology}
One of the standard ways to quantify medium effects on particle production is the nuclear modification factor, $R_{AA}$. This observable is defined as the ratio between the particle yield in nucleus-nucleus collisions with mass number $A$, $N^{AA}$, and the yield in proton-proton collisions, $N^{pp}$, normalized by the number of binary nucleon-nucleon collisions $N_{coll}$ expected in the nuclear interaction:
\begin{equation}
  R_{AA} =\frac{N^{AA}}{N^{pp}N_{\text{coll}}}~.\label{RAA}
\end{equation}
In the absence of nuclear effects, $N^{AA}=N^{pp}N_{\text{coll}}$ and the nuclear modification factor equals 1 by construction.

When initial cold nuclear matter (CNM) effects are taken into account, the nuclear yields are modified. Among these, the shadowing of the nuclear parton distribution functions (nPDFs) is the dominant contribution, see details of our implementation in~\cite{Escobedo:2021ttu,Armesto:2025lbd}. In this case, one can write:
\begin{equation}
    N^{AA}_{\text{CNM}}=N_{\rm coll}S^{\rm sh}N^{pp}~,
\end{equation}
where $S^{\rm sh}$ corresponds to the shadowing factor.

In the presence of the medium the yield gets modified due to suppression and coalescence effects. Recalling the notation for the survival probability from~\eqref{eq:survival} and the regeneration factor from~\eqref{eq:R}, the full nuclear yield for one particle species is taken to be
\begin{equation}
  N^{AA} =    N^{AA}_{\text{CNM}}S(t_0,t_f)\Theta(T_d - T_0)+\int_{t_0}^{t_f} dt R(t)S(t,t_f)\Theta(T_d - T)~,\label{fullYield}
\end{equation}
where $\Theta(x)$ is the Heaviside step function and $T_d$ is the dissociation temperature of the species. 

In Eq.~\eqref{fullYield}, two distinct contributions can be identified: the initially produced particles, which are suppressed from the initial time $t_0$ up to the freeze-out time $t_f$, and the regenerated particles, which are suppressed from their formation time $t > t_0$ until $t_f$. The overall suppression in Eq.~\eqref{fullYield} is implemented through two mechanisms: the survival probability, which accounts for the effect of the imaginary part of the potential (i.e., Landau damping), and a Heaviside step function, introduced to model a sharp Debye screening at temperatures above $T_d$.

The full nuclear modification factor can be calculated by introducing Eq.~\eqref{fullYield} into Eq.~\eqref{RAA}. In the next subsections it will be computed for different quarkonia states.

In a realistic heavy-ion collision the temperature is not spatially uniform.  We
account for this by applying the survival probability locally at each transverse
position $\mathbf{x}_\perp$, using the initial temperature profile
$T_0(\mathbf{x}_\perp)$ from the model of
Ref.~\cite{Escobedo:2021ttu}.

\subsection{Phenomenological inputs}

The following ingredients are needed to evaluate the nuclear modification
factor:

\paragraph{Medium evolution}
We use Bjorken expansion with initial conditions $T_0 = 500\;\text{MeV}$ and
$t_0 = 0.6\;\text{fm}$, appropriate for Pb--Pb collisions at $\sqrtsNN = 5.02\;\text{TeV}$.
The transverse temperature profile and the number of binary collisions $N_\text{coll}$
are obtained from a Glauber model supplemented with the shadowing model of
Ref.~\cite{Escobedo:2021ttu}.

\paragraph{Initial open-charm and open-bottom production}
The initial number of free charm (bottom) quarks is estimated from binary scaling,
$N_{Q\bar{Q}}^{(0)} = N_\text{coll}\,\sigma_{pp\to Q\bar{Q}}/\sigma_{pp}$.
We use the ALICE measurements $d\sigma_{pp\to c\bar{c}}/dy\big|_{|y|<0.5} = 1.165\;\text{mb}$ and 
$d\sigma_{pp\to b\bar{b}}/dy\big|_{|y|<0.5} = 34.5\ \mathrm{\mu b}$ 
at $\sqrtsNN = 5.02\;\text{TeV}$~\cite{ALICE:2021dhb,ALICE:2021mgk}.

\paragraph{pp baseline cross sections}
For $\jpsi$ we use the ALICE midrapidity cross section
$d\sigma_{\jpsi}^{pp}/dy\big|_{y=0}=5.64\ \mathrm{\mu b}$ at $\sqrtsNN = 5.02\;\text{TeV}$~\cite{Acharya:2019lkh}, including prompt production only. 
For the $\ups$ we use the CMS measurement at
$\sqrtsNN = 5.02\;\text{TeV}$~\cite{Sirunyan:2018nsz},
$d\sigma_{\Upsilon(1S)}^{pp}/dy\big|_{y=0}=60.2$ nb.

\paragraph{Cold nuclear matter effects}
CNM effects are modeled as shadowing using the framework in
Ref.~\cite{Escobedo:2021ttu}.

\subsection{Results for $\jpsi$}

\begin{figure}[h]
\centering
\includegraphics[scale=0.45]{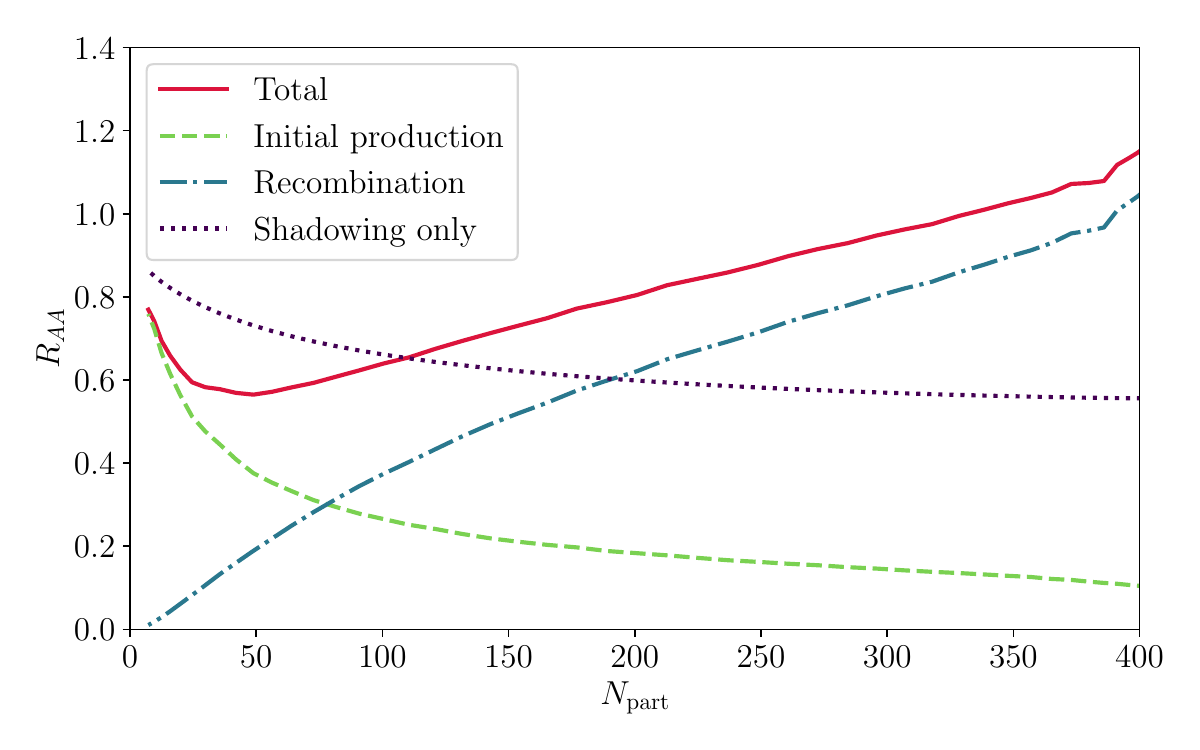}
\caption{Nuclear modification factor $\raa$ of the $\jpsi$ in Pb--Pb collisions
at $\sqrtsNN = 5.02\;\text{TeV}$.  The dotted curve shows only cold nuclear matter
effects (shadowing); the dashed curve adds suppression from the complex Gauss-law
potential; the dash-dot curve shows the recombination contribution alone; and the
solid curve gives the total result.}
\label{fig:raa_jpsi}
\end{figure}

Figure~\ref{fig:raa_jpsi} shows our prediction for $\raa^{\jpsi}$ in Pb--Pb
collisions at $\sqrtsNN = 5.02\;\text{TeV}$ as a function of $N_\text{part}$,
including both suppression and recombination contributions.

The $\jpsi$ displays a sizeable recombination contribution in central collisions
($N_\text{part} \gtrsim 200$), driven by the relatively large charm-quark
multiplicity at LHC energies.  This recombination partially compensates the
suppression, leading to a total $\raa$ around unity, significantly larger than the suppression-only result. The $\pT$-integrated
prediction is dominated by low-$\pT$ coalescence, consistent with the well-known
regeneration picture for $\jpsi$ at the LHC~\cite{Braun-Munzinger:2000px,Thews:2000rj}.

\subsection{Results for $\ups$}

\begin{figure}[h!]
\centering
\includegraphics[scale=0.45]{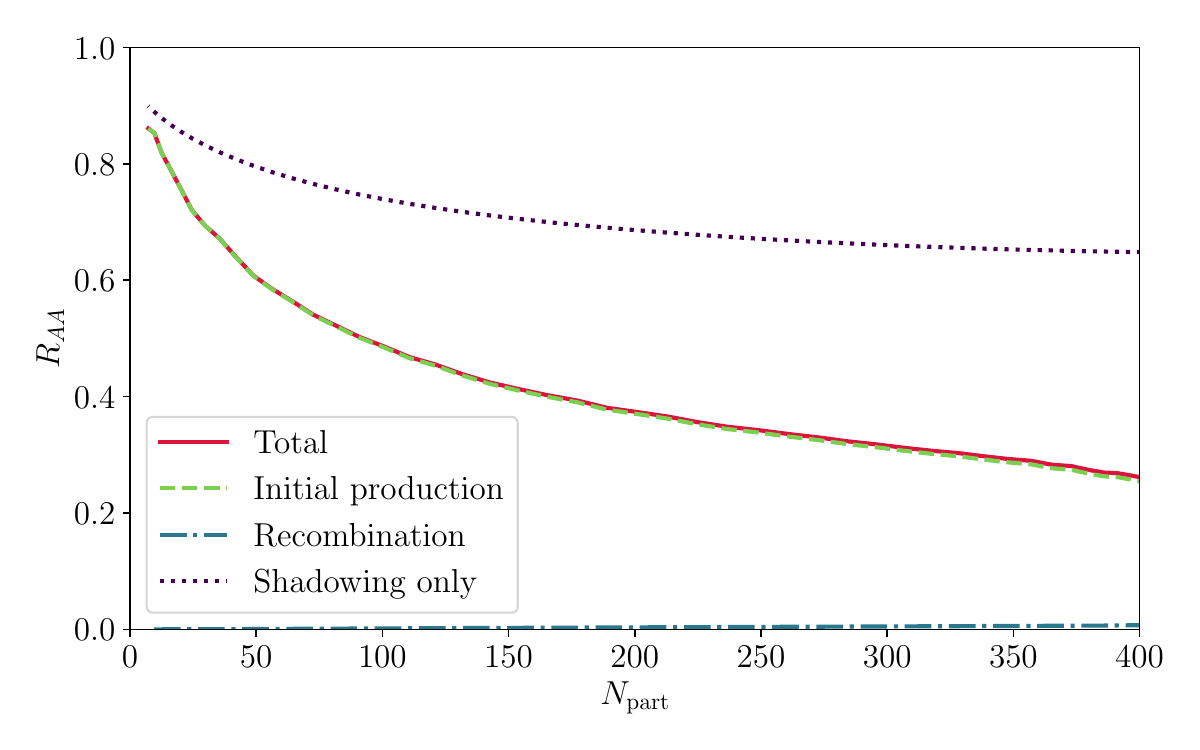}
\caption{Nuclear modification factor $\raa$ of the $\ups$ in Pb--Pb collisions
at $\sqrtsNN = 5.02\;\text{TeV}$.  Curves and symbols as in
Fig.~\ref{fig:raa_jpsi}.  The recombination contribution is significantly
smaller than for the $\jpsi$ due to the much smaller bottom-quark multiplicity, and it becomes barely visible.}
\label{fig:raa_upsi}
\end{figure}

Figure~\ref{fig:raa_upsi} shows the corresponding prediction for $\raa^{\ups}$.
In contrast to the $\jpsi$ case, the recombination contribution to the $\ups$ is
strongly suppressed.  The bottom-quark multiplicity at LHC energies is roughly
two orders of magnitude smaller than the charm-quark multiplicity, so the
$N^2/V$ factor in Eq.~\eqref{eq:R1S} is dramatically reduced.  The total $\raa$
of the $\ups$ is therefore dominated by suppression, with only a minor correction
from coalescence.

\subsection{Comparison to experimental data}

\begin{figure}[h!]
\centering
\includegraphics[scale=0.45]{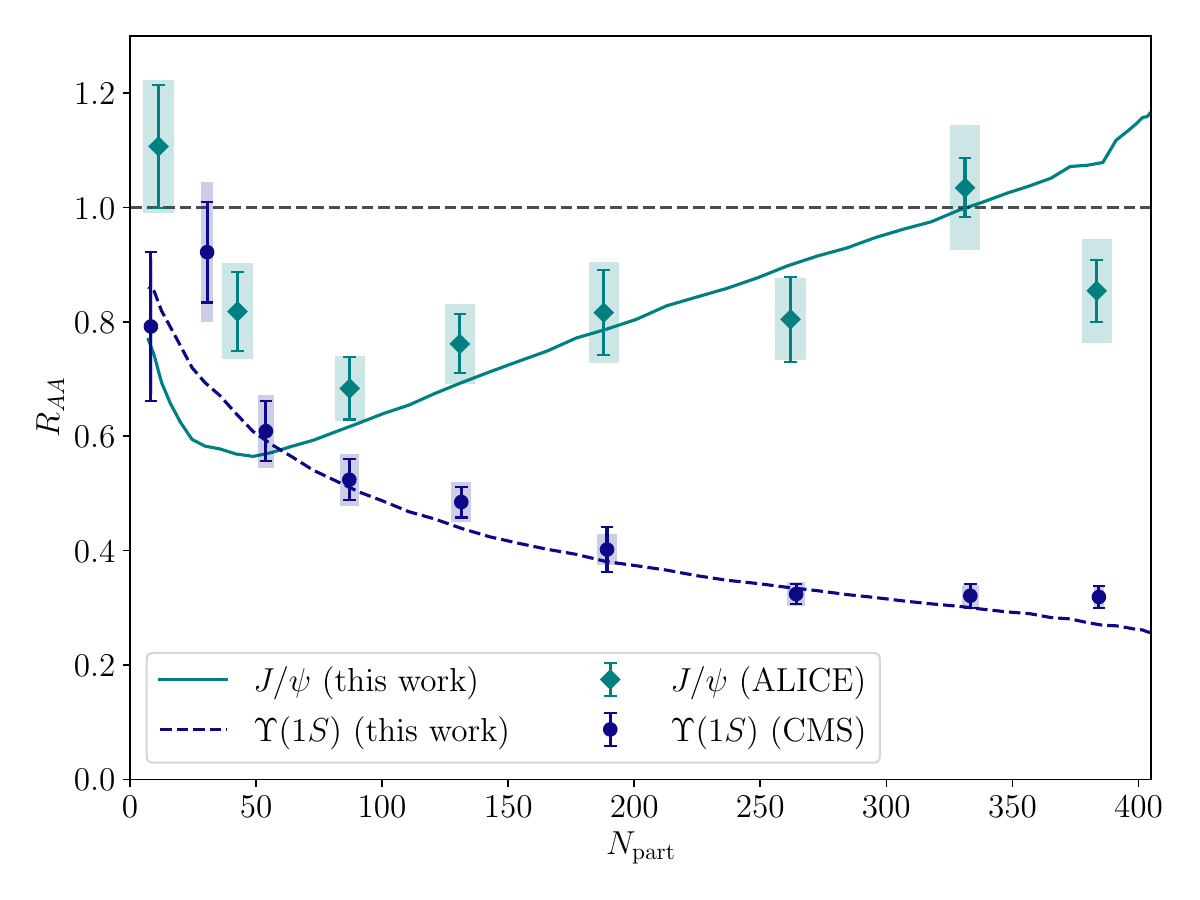}
\caption{Nuclear modification factor $\raa$ of the $\jpsi$ and $\ups$ in Pb--Pb collisions
at $\sqrtsNN = 5.02\;\text{TeV}$. 
Experimental data from ALICE~\cite{Acharya:2019lkh} and 
CMS~\cite{Sirunyan:2018nsz,Sirunyan:2018nsz_jpsi} are shown for comparison.}.
\label{fig:comparison}
\end{figure}

In Figure~\ref{fig:comparison}, we show our results for the nuclear modification factors of both $\jpsi$ and $\ups$ compared to the experimental data available from ALICE~\cite{Acharya:2019lkh} and CMS~\cite{{Sirunyan:2018nsz,Sirunyan:2018nsz_jpsi}}. The agreement is reasonable, considering the lack of fitting parameters in our approach where the relaxation times are taken from the literature~\cite{Song:2012at,Wu:2024gil} for $J/\psi$ and extrapolated for the $\Upsilon$ following the mass difference.

\section{Summary and conclusions} 
\label{sec:summary}

We have applied the Lindblad-driven open quantum system framework to the
production of the conventional quarkonium ground states $\jpsi$ and $\ups$ in
Pb-Pb collisions at $\sqrtsNN = 5.02\;\text{TeV}$.  Our approach proceeds as follows:
First, we use the Gauss-law model of
    Lafferty and Rothkopf~\cite{Lafferty:2019jpr}, which provides
    a single-parameter ($\mD(T)$) parametrization of both $\ReV$ and $\ImV$, the complex in-medium potential.
By solving the Schrödinger equation with the complex potential and extracting the imaginary part of the corresponding Green's function, we obtain the quarkonium spectral functions for both charmonium and bottomonium across a wide range of temperatures. The spectral shape directly encodes the in-medium properties of each state: the peak positions, after subtracting the masses of the two constituent quarks, yield the binding energies, the half-widths at half-maximum give the thermal decay widths, and the temperature at which a given peak merges into the continuum defines the dissociation temperature. From this analysis we find $T_{J/\psi} \approx 372$ MeV and $T_{\psi(2S)} \lesssim 155$ MeV for charmonium, and $T_{\Upsilon(1S)} \approx 564$ MeV, $T_{\Upsilon(2S)} \approx 223$ MeV, $T_{\Upsilon(3S)} \approx 164$ MeV and $T_{\Upsilon(4S)} < 150~\text{MeV}$ for bottomonium, with the thermal decay widths entering directly the survival probability under Bjorken expansion.

Following Ref.~\cite{Armesto:2025lbd}, we derive a coalescence model directly from the Lindblad equation by projecting the stochastic jump operators onto the bound-state subspace. Under the adiabatic approximation, the recombination rate $R_{1S}(t)$ is proportional to $N^2/V$, where $N$ is the number of free heavy quarks in the medium volume $V$, and is computed using the vacuum wave functions obtained from the Schrödinger equation together with the fluctuation-dissipation theorem, which identifies the normalization factor $Z_{t,1S}$ with the thermal decay width $\Gamma_{1S}(T)$. We further correct the number of heavy quark-antiquark pairs available for recombination in the medium by a modulation due to the finite relaxation time of heavy quarks. The time-dependent recombination yield is then given by Eq.~(\ref{eq:R}), where each regenerated state is subsequently suppressed by the survival probability from its formation time until freeze-out. At LHC energies, the charm-quark multiplicity is large enough to make recombination a significant contribution for the $J/\psi$ in central collisions, partially compensating the suppression and bringing the total $R_{AA}$ into agreement with ALICE and CMS measurements. In contrast, the bottom-quark multiplicity is roughly two orders of magnitude smaller, so the $N^2/V$ factor is dramatically reduced and recombination remains negligible for the $\Upsilon(1S)$, whose $R_{AA}$ is therefore governed almost entirely by suppression.

The Lindblad framework provides a unified, first-principles-inspired treatment of both suppression and recombination, with a direct microscopic link between the complex potential and the observable yield. Unlike phenomenological approaches in which suppression and regeneration are modeled independently and then combined by hand, the present framework derives both mechanisms from a single equation of motion for the density matrix of the heavy quark–antiquark pair. The imaginary part of the in-medium potential enters directly as the generator of decoherence and decay, determining the thermal decay widths and hence the survival probability, while the same jump operators that drive dissociation also govern the recombination rate. This internal consistency ensures that suppression and regeneration are not treated as competing corrections but as two complementary manifestations of the same underlying dynamics. The resulting description is therefore not only more theoretically grounded than traditional transport models, but also more predictive: once the complex potential is fixed, here through the Gauss-law model, all medium effects on the quarkonium yield follow without additional free parameters. We regard this as a significant step toward a truly first-principles description of quarkonium production in heavy-ion collisions.

\bigskip
\paragraph{Acknowledgements}
NA, EGF and VLP are supported by European Research Council project ERC-2018-ADG-835105 YoctoLHC, by Xunta de Galicia (CIGUS Network of Research Centres), by European Union ERDF, and by the Spanish Research State Agency under projects PID20231527\-62NB---I00 and CEX2023-001318-M financed by MICIU/AEI/10.13039/501100011033. The work of MAE has been supported by the Maria de Maeztu excellence program under project CEX2024-001451-M, and by project PID2022-136224NB-C21 funded by MICIU/AEI/10.13039/501100011033, and by grant 2021-SGR-249 of Generalitat de Catalunya. MAE acknowledges the hospitality of the MITP during the workshop Exotic Quarkonia in Heavy-ion Collisions and the discussions with its participants. VLP has been supported by Xunta de Galicia under project ED481A2022/286.

\bibliographystyle{unsrtnat}
\bibliography{quarkonium_lindblad}

\end{document}